\documentstyle[preprint,revtex]{aps}
\begin{document}
\draft
\begin{title}
High energy photon-nucleon and photon-nucleus cross sections
\end{title}
\author{Katsuhiko Honjo\cite{kh} and Loyal Durand\cite{ld}}
\begin{instit}
Department
of Physics, University of Wisconsin--Madison,\\
Madison, Wisconsin
53706
\end{instit}
\author{Raj Gandhi\cite{rg}}
\begin{instit}
Department of Physics, Texas A\&M University\\
College Station, TX 77843
\end{instit}
\author{Hong Pi}
\begin{instit}
Department of Theoretical Physics, University of Lund\\
Lund, Sweden S-22362
\end{instit}
\author{Ina Sarcevic\cite{is}}
\begin{instit}
Department of Physics, University of Arizona,\\
Tucson, Arizona 85721
\end{instit}
\begin{abstract}
We re\,examine the theory of hadronic photon-nucleon
interactions at the quark-gluon level. The possibility of multiple
parton collisions in a single photon-nucleon collision requires an
eikonal treatment of the high-energy scattering process.
We give a general formulation of the theory in which the $\gamma p$
cross section is expressed as a sum over properly eikonalized cross
sections for the interaction of the virtual hadronic components of the
photon with the proton, with each cross section weighted by the
probability with which that component appears in the photon, and then
develop a detailed model which includes contributions from light
vector mesons and from excited virtual states described in a
quark-gluon basis.  The parton distribution functions which appear can
be related approximately to those in the pion, while a weighted sum
gives the distribution functions for the photon.  We use the model to
make improved QCD-based
predictions for the total inelastic photon-nucleon and
photon-nucleus cross sections at energies relevant for  HERA
experiments and cosmic ray observations.
We emphasize the importance in this procedure of including
a soft-scattering background such that the
calculated cross sections join smoothly with
low-energy data.
Our results show clearly that high-energy
measurements of the total inelastic $\gamma p$ cross section can
impose strong constraints on the gluon and quark distributions in
the photon, and indirectly on those in the pion.

\end{abstract}

\section{INTRODUCTION}

The high-energy ``hadronic" interactions of the photon are
currently a subject of considerable experimental and theoretical
interest. The virtual quark-gluon structure of the photon can
lead to a significant inclusive jet cross section in high-energy
$\gamma$-$p$ collisions. This cross section and the quark and gluon
distributions in the proton will be measurable in future
experiments at the electron-proton collider HERA at DESY
\cite{1,2,3}. It is useful in that context to have predictions
\cite{3} for
the total inelastic $\gamma$-$p$ cross section through the HERA
energy range. The hadronic interactions of the photon in
very-high-energy photon-nucleon and photon-nucleus collisions have
also been studied recently \cite{4,5,6,7} in the context of reports
of anomalously high muon content in PeV cosmic-ray air showers
\cite{8}. These showers were apparently initiated by photons from
point sources of cosmic rays. However, their reported muon content
was up to an order of magnitude higher than expected in
electromagnetic showers, and approached that typical of
proton-initiated hadronic showers. This suggested that the hadronic
structure of the photon might play a role \cite{4}, an idea which
has been investigated in detail \cite{5,6,7}.

Although it now appears that ``hadronic" photon-nucleus
interactions with realistic photon structure functions cannot
account for the apparent muon anomalies \cite{6}---and the
cosmic-ray observations have so far not been confirmed---theoretical
questions with respect to the proper calculation of
photon-nucleon and photon-nucleus
cross sections remain. In particular, the large numbers of partons
(quarks and gluons) present in the photon and nucleons at very high
energies make multiple parton-parton collisions likely in a single
$\gamma$-nucleon interaction, and can lead to violations of
partial-wave unitarity in simplified calculations. An eikonal treatment
of the scattering amplitude is necessary to remove this problem.
It was pointed out recently by Collins and Ladinsky \cite{9} that the
method of eikonalization used in previous work \cite{6} was
incorrect. Those calculations treated the hadronic photon similarly
to any other hadron, and used the formalism developed for
nucleon-nucleon scattering \cite{10} to describe the photon-nucleon
interaction. However, the photon is not like an ordinary hadron
which always exists in a hadronic state consisting of quarks and
gluons. The photon is usually in a bare or
non-hadronic state with respect to the strong interactions and
only turns into a virtual $q\bar q$ pair occasionally, with a total
probability ${\cal P}_{\rm had}\ll 1$. The eikonalization should
apply only to the scattering of the resulting hadronic system.
The probability that the photon is in any particular hadronic state
should therefore appear as an overall factor in the contribution of
that state to $\sigma_{\gamma p}$, and must be extracted  from
the corresponding eikonal function.

Collins and Ladinsky \cite{9} estimated ${\cal P}_{\rm had}$ using
the vector dominance model with only the lightest vector states
included, with the result
\begin{equation}
{\cal P}_{\rm
had}^{CL}=\frac{4\pi\alpha_{em}}{f_\rho^2}\approx\frac{1}{300}.
\label{eq2}
\end{equation}
It was assumed that this factor could simply be divided out of
the QCD contributions to the eikonal function as defined in
earlier work, and included as an overall factor in
$\sigma^{\gamma p}$ as noted above.
Following this line of reasoning, Collins and Ladinsky \cite{9},
Fletcher, Gaisser and Halzen \cite{11}, and Forshaw and Storrow
\cite{12} re-computed the $\gamma p$ total inelastic cross section
up to cosmic-ray energies and found it to be considerably smaller
than previously suggested \cite{4,5,6}.

We have also re\,examined
this question and have improved earlier predictions \cite{3} for the
photon-nucleon and photon-nucleus cross sections at energies
relevant to  projects at HERA, and
also the cosmic-ray observations \cite{6}.
In the process, we have considerably improved
the theoretical treatment of the problem \cite{13}.  The $\gamma p$
cross section is now expressed as a weighted sum of eikonalized
cross sections for the interactions of the hadronic components of
the photon with the proton.  The weights are the probablilities
with which those virtual components appear in the physical photon.
We have developed a detailed model which includes contributions
from light vector mesons and from excited virtual states with the
latter described in a quark-gluon basis. The ``soft" contributions
to the eikonal functions are parametrized in the form expected from
Regge theory to assure that the calculated high-energy cross
sections connect smoothly with the cross sections measured
at lower
energies.  The parton distribution functions which appear in the
``hard scattering" contributions to the eikonal functions can be
related approximately to those in the pion.  This approach allows
us to use the (not-well-determined) pion distributions to predict
the high-energy behavior of the $\gamma p$ cross section, and also
gives an approximate prediction for the parton distributions in the
photon  in terms of these in the pion.  We also
develop an alternative approach based directly on the photon
distribution functions, but find that the simple approximations
used in \cite{9,11,12} must be modified.

We find, in accord with the results of other
authors \cite{9,11,12} that the inelastic $\gamma p$ cross
section is strongly suppressed at high energies relative to
results reported earlier \cite{4,5,6,7}.  The new results on
$\gamma$-air interactions make it quite clear that the hadronic
interactions of the photon cannot explain the reported muon
anomalies in cosmic-ray showers, if the anomalies in fact exist.
However, the QCD contributions to the hadronic interactions of
the photon still lead to a rapid rise in $\sigma^{\gamma p}$ at
HERA energies as predicted in earlier calculations \cite{3} and
observed in recent experiments \cite{30}. The magnitude of the rise provides
a quantitative test of the whole picture.
In particular, our results show clearly that measurements of the
total inelastic $\gamma p$ cross section at HERA can impose strong
constraints on the parton distributions in the photon, and
indirectly on those in the pion.

In the following sections, we sketch the theoretical background
of our calculations (Sec.\ II), discuss the
detailed model
(Sec.\ III), and present our
calculations and
conclusions (Sec.\ IV).

\section{THEORETICAL BACKGROUND}

\subsection{The physical photon state and the $\gamma p$ cross
section}

As noted above, the hadronic interactions of the photon are
different from those of ordinary hadrons in that the photon exists
in a hadronic state only with a small total
probability ${\cal P}_{\rm
had}$. In the presence of the strong interactions---but neglecting
purely electromagnetic effects---the incoming physical photon in a
$\gamma p$ scattering process can be described perturbatively as
a superposition of a ``bare" photon and a set of virtual hadronic
states,
\begin{equation}
|\gamma\rangle_{\rm phys}=(1-{\cal P}_{\rm
had})^{1/2}|\gamma\rangle_{\rm bare}+
\sum_m\mbox{}' |m\rangle\frac{\langle m|H_I|\gamma\rangle}{%
E_\gamma-E_m},
\label{eq4}
\end{equation}
where $H_I=H_I({\bf x},0)$ is the $\gamma q\bar q$ interaction
\begin{equation}
H_I=j\cdot A=e\sum_q e_q \bar\psi_q \not\!\! A \psi_q ({\bf x},0).
\end{equation}
The sum is restricted to hadronic states which
can be treated as real incoming states in a
subsequent scattering, that is, to states which last a time
long relative to the time necessary for the photon to traverse the
target proton or the relevant structures in it, hence are almost
on-shell.
The ``bare" state $|\gamma\rangle_{\rm bare}$ includes the short-lived
hadronic fluctuations which cannot be distinguished in the $\gamma p$
interaction.
The probability with which an observable hadronic state $|m\rangle$
appears in $|\gamma\rangle_{\rm phys}$ is simply
\begin{equation}
{\cal P}_m = \frac{\left| \langle m | H_I | \gamma \rangle \right|^2}%
{\left( E_\gamma - E_m \right)^2},
\end{equation}
and
\begin{equation}
{\cal P}_{\rm had} = \sum_m\mbox{}' {\cal P}_m.
\end{equation}

Given this structure, the inelastic $\gamma p$ cross section will be
given approximately as a sum of cross sections,
\begin{equation}
\sigma_{\rm inel}^{\gamma p} = (1 - {\cal P}_{\rm had})\sigma_{\rm dir}
+\sum_m\mbox{}' {\cal P}_m \sigma_m.
\end{equation}
The first term represents the inelastic interaction of the bare photon
directly with the proton (or the quarks and gluons in it); the sum
includes the interactions of the quasi-real intermediate hadronic
states with the proton. We neglect possible interference terms
involving final states which can be reached from distinct initial
configurations. We expect these to give very small average contributions.

There are no rescattering corrections to the direct-interaction cross
section $\sigma_{\rm dir}$ to lowest order in the electromagnetic
coupling $\alpha_{\rm em}$, and $\sigma_{\rm dir}$ can simply be
identified at high energies and momentum transfers with the inclusive
direct-interaction cross section of QCD,
\begin{equation}
\sigma_{\rm dir}^{QCD}=\sum_i\int_0^1dx
\int_{p^2_{\perp,\rm min}} dp_\perp^2\,
f_i^p(x,p_\perp^2)\,\frac{d\hat\sigma_{\gamma i}}{dp_\perp^2}
(\hat s, p_\perp^2).
\label{eq8}
\end{equation}
Here the $f$'s are the number distributions of the quarks,
antiquarks, and gluons in the proton, $d\hat\sigma_{\gamma i}/
d p_\perp^2$ is the relevant differential $\gamma$-parton
scattering cross section, and $\hat s$ is the invariant mass of the
$\gamma$-parton system.
The lower bound $p_{\perp,\rm min}$ on the transverse momentum
$p_\perp$ in the scattering parametrizes the point at which
semihard and soft processes merge.

The theoretical problem at this point involves the calculation of the
probabilities ${\cal P}_m$ and the hadronic cross sections $\sigma_m$
in Eq.\ (6).  The low-mass hadronic states in the sum are just the
vector meson states $\rho$, $\omega$, $\phi$. These are normal hadronic
states produced with probabilities \cite{14}
\begin{equation}
{\cal P}_V=\frac{4\pi\alpha_{\rm em}}{f_V^2},
\end{equation}
where $f_V$ is the $\gamma V$ coupling in the effective Hamiltonian of
the vector dominance model,
\begin{equation}
H_{\gamma V} = \frac{e}{f_V} m_V^2 {\bf A \cdot V}\ .
\end{equation}
The scattering of the vector mesons can be described using the eikonal
methods developed for high-energy nucleon-nucleon \cite{10} or
meson-nucleon \cite{15} scattering as discussed below. We will
treat the $\rho$ and $\omega$ as
equivalent, and will use the quark-model ratios for the
couplings
$f_V$. The contribution of the low-mass vector mesons to the sum in
Eq.\ (6) is then
\begin{equation}
\sum_V {\cal P}_V \sigma_V =
\lambda {\cal P}_\rho \sigma_\rho = \lambda
\frac{4\pi \alpha_{\rm em}}{f_\rho^2} \sigma_\rho,
\end{equation}
where $\sigma_\rho$
is the inelastic $\rho p$ scattering cross section, $\lambda=4/3$
for equal $\rho p$, $\omega p$, and $\phi p$ cross sections, and
$\lambda=10/9$ for complete suppression of the $\phi$ contribution
\cite{16}.

There is no strong resonance structure observed in the vector $u\bar
u$, $d\bar d$, and $s\bar s$ channels at higher masses, but many
possible nonresonant final states exist in inelastic $\gamma p$
scattering. In this situation the relevant hadronic states in Eq.\
(6) are best described inclusively in a quark-gluon basis. We begin
by considering the initial transition $\gamma \rightarrow q\bar q$.
We write the 4-momenta $k$, $p$, and $p'$ of the photon, quark and
antiquark in terms of longitudinal and transverse components as
\begin{eqnarray}
k &=& (k,{\bf 0}_\pm,k),\nonumber\\
p &=& \left( \sqrt{(xk)^2+p_\perp^2}, {\bf p}_\perp,
xk\right),\nonumber\\
p' &=&  \left( \sqrt{(1-x)^2 k^2 + p_\perp^2},-{\bf p}_\perp,
(1-x)k \right),
\end{eqnarray}
where ${\bf k\cdot p_\perp} = 0$ and $x={\bf p}\cdot \hat k/
|{\bf k}|$ is the fraction of the (purely longitudinal) 3-momentum of
the photon which is carried by the quark. The differential transition
probability $d{\cal P}_{q\bar q}$
for $\gamma \rightarrow q\bar q$
is then given for $xk$, $(1-x)k \gg
p_\perp$ and one flavor of quark by
\begin{equation}
d{\cal P}_{q\bar q} = \frac{3\alpha_{\rm em} e_q^2}{2\pi}
\frac{1}{p_\perp^2}\left[ x^2 + (1-x)^2 \right] dx dp_\perp^2.
\end{equation}
With the same kinematic approximations, $p_\perp$ and the invariant
mass $M$ of the $q\bar q$ pair are related by
\begin{equation}
M^2 = (E + E')^2 - {\bf k}^2 = p_\perp^2/x(1-x) \ll {\bf k}^2\ .
\end{equation}

The lifetime of a $q\bar q$ system with mass $M$ is given approximately
by
\begin{equation}
\tau \approx 1/(E+E'-E_\gamma) \approx 2k/M^2\ .
\end{equation}
The average longitudinal separation of the quarks during the lifetime
of the system,
\begin{eqnarray}
r_\parallel & \approx & \frac{1}{2}\tau v_{\parallel,\rm relative}
\nonumber\\
& \approx & \frac{1}{2}\tau \left(
\frac{xk}{E} - \frac{(1-x)k}{E'} \right) \approx
\frac{1}{2k} \frac{2x-1}{x(1-x)}\ ,
\end{eqnarray}
is much smaller than the
average transverse separation
\begin{equation}
r_\perp = \frac{1}{2}\tau\left(\frac{p_\perp}{E} -
\frac{(-p_\perp)}{E'}\right) \approx \frac{1}{p_\perp}
\end{equation}
for $p_\perp \ll k$.
If $r_\perp$ is greater than, or on the order of, the average
transverse radius $R_\perp$ of a vector meson, QCD confinement effects
will clearly set in, and the $q\bar q$ system will appear in a hadronic
collision as a ``soft" system with a typical hadronic size and
interactions, e.g., as a light vector meson. On the other hand, for
$r_\perp < R_\perp$, the system of quark, antiquark, and the
connecting color flux tube will be smaller than a vector meson during
its transit through the target. It may still interact softly with the
target, but with an
intrinsic cross section which is reduced by a geometrical
factor of order $(r_\perp/R_\perp)^2 \approx (1/p_\perp R_\perp)^2$
for large initial values of $p_\perp$. The high momentum transfer
branching processes of perturbative QCD lead to higher transverse
velocities but shorter intermediate lifetimes, and to only a small
average increase in the size of the system.

We can estimate $R_\perp$ using
 the expected similarity of the $\rho$ and $\pi$
meson wave functions
and measurements of the pion form factor
\cite{17}.
The measured parameter
$\mu^2=0.47\rm\ (GeV/c)^2$ \cite{17}
in the series representation of
the pion electromagnetic form factor
\begin{equation}
G_\pi(Q^2)=1-Q^2/\mu^2 + \cdots
\end{equation}
corresponds to a root-mean-square transverse radius
\begin{equation}
\langle r_\perp^2
\rangle_\pi^{1/2} = 2/\mu \approx (350\rm\ MeV)^{-1}.
\end{equation}
We will identify $R_\perp$ with this radius. Under this
assumption,
the vector-meson-like behavior will hold for initial values of
$p_\perp$ less than about 350 MeV.
This value corresponds to an average mass
of the system
$\langle M \rangle = \frac{9\pi}{8} p_\perp \approx 1.2$
GeV, where the average is over the
$x$ distribution in Eq.\ (12). This
mass is comfortably
between the masses of the $\rho(770)$
and $\omega(783)$, which we are
taking into account explicitly, and the $\rho(1450)$, which we are
not.
However, it is sufficiently close to the mass of the $\phi(1020)$
that $\phi$ production, which involves the relatively massive
strange quarks, could be suppressed, a possibility which we will
consider later.
In the following discussion
we will adopt the value $Q_0 = \mu/2 \approx 350$ MeV/c for the
critical value of $p_\perp$ above which the hadronic states are no
longer describable as vector mesons.
The precise value of this parameter will not be essential.

Our expression for the inelastic $\gamma p$ cross section is given at
this point by
\begin{eqnarray}
\sigma_{\rm inel}^{\gamma p} &=& (1-{\cal P}_{\rm had})\sigma_{\rm dir}
+\lambda{\cal P}_\rho \sigma_\rho\nonumber\\
&& + \int_{Q_0^2} d{\cal P}_{q\bar q}(x_0,p_{\perp 0}^2)
\sigma_{q\bar q},
\end{eqnarray}
where ${\cal P}_{\rm had}$ may be dropped in the first term to leading
order in $\alpha_{\rm em}$. The integral in the last term is over the
initial values of $x$ and $p_\perp$ for the $q\bar q$ pair. The
condition that the intermediate hadronic state have a lifetime long
compared to the time necessary to traverse the target imposes an upper
limit on the integrations given by
\begin{equation}
\tau \approx 2k/M^2 = 2k x(1-x)/p_\perp^2 \gg R\ ,
\end{equation}
where $R$ is the radius of the target. Thus,
\begin{equation}
p_\perp^2 \ll 2kx(1-x)/R\leq k/2R.
\end{equation}

This limit has little effect at high energies because of the
$1/p_\perp^2$ decrease of $\sigma_{q\bar q}$ at large momentum
transfers, and we will ignore it. For simplicity, we will also ignore
the possible dependence of $\sigma_{q\bar q}$ on $x_0$. Then, using the
expression in Eq.\ (12) and integrating over $x_0$, we find that, to
leading order in $\alpha_{\rm em}$ \cite{new18}
\begin{eqnarray}
\sigma_{\rm inel}^{\gamma p} &=& \sigma^{\rm QCD}_{\rm dir} +
\lambda {\cal P}_\rho \sigma_\rho\nonumber\\
&& + \sum_q e_q^2 \frac{\alpha_{\rm em}}{\pi}
\int_{Q_0^2} \frac{d p_{\perp 0}^2}{p_{\perp 0}^2}
\sigma_{q\bar q} (s, p_{\perp 0}^2).
\end{eqnarray}
In the next section, we discuss the calculation of the hadronic cross
sections $\sigma_\rho$ and $\sigma_{q\bar q}$.

\section{Estimates for the hadronic cross sections}

\subsection{Vector meson cross sections}

The $\rho p$ cross section and the lower-mass $q\bar q p$ cross
sections in Eq.\ (22) are of hadronic magnitude, and are best treated
at high energies using an eikonal formalism. We will follow the
methods discussed in detail in \cite{10} and \cite{15}, and write a
typical hadronic cross section in Eq.\ (22) as
\begin{equation}
\sigma_m(s) = \int d^2 b \left(
1 - e^{\textstyle-2{\rm Re}\chi_m(b,s)}\right).
\end{equation}
Semiclassically, $e^{-2{\rm Re}\chi}$ is the probability that the two
incoming systems (the proton and the virtual hadronic component
$|m\rangle$ of the
photon) survive a collision at impact parameter $b$. We will write
the eikonal function $\chi$ as the sum of two terms corresponding to
contributions from ``soft" and ``hard" or low- and
high-momentum-transfer processes, \begin{equation}
{\rm Re}\chi = {\rm Re}\chi_{\rm soft} + {\rm Re}\chi_{\rm QCD}.
\end{equation}

In the case of the $\rho$ meson, we are dealing with a normal hadron,
and familiar arguments \cite{10} can be used to relate the hard
component of $\chi$ to
$\sigma_{\rm QCD}^{\rho p}$,
the inclusive parton-level cross section for $\rho p$
scattering,
\begin{equation}
2{\rm Re} \chi_{\rm QCD}^{\rho p} (b,s) =
A_{\rho p}(b) \sigma_{\rm QCD}^{\rho p}
\end{equation}
where
\begin{equation}
\sigma_{\rm QCD}^{\rho p}
=
 \sum_{ij}
\frac{1}{1+\delta ij}
\int_0^1 dx_1 \int_0^1 dx_2 \int_{p_\perp^2,
\rm min} dp_\perp^2
 f_i^p (x_1, p_\perp^2) f_j^\rho (x_2, p_\perp^2)
\frac{d\hat\sigma_{ij}}{d p_\perp^2}\ .
\end{equation}
Here $f^p$ and $f^\rho$ give the parton distributions in the proton and
$\rho$ meson, and $d\hat\sigma_{ij}/d p_\perp^2$ is the cross section
for the scattering of partons $i$ and $j$. The proton structure
functions are well known \cite{18}. We will return later to a
discussion of the structure functions for the $\rho$ meson.
$A_{\rho p}(b)$ is the parton density overlap function, expressed as
the convolution of the normalized transverse parton densities in the
$\rho$ meson and the proton,
\begin{equation}
A_{\rho p}(b) = \int d^2 b' \rho_\rho (b) \rho_p
\left( \left|\vec b - \vec b'\right|\right),\qquad
\int d^2 b A_{\rho p}(b) = 1.
\end{equation}

The density $\rho_p(b)$ can be taken as the Fourier transform of the
proton electromagnetic charge form factor,
\begin{equation}
G_E\left( k_\perp^2\right) \approx \left( 1 +
\frac{k_\perp^2}{\nu^2} \right)^{-2},\qquad
\nu^2 = 0.71\rm\ GeV^2,
\end{equation}
a choice which works well in the case of $pp$ and $\bar pp$ scattering
\cite{10}. Then
\begin{eqnarray}
\rho_p(b) &=& \frac{1}{(2\pi)^2} \int d^2 k_\perp G_p
\left( k_\perp^2 \right) e^{i{\bf k}_\perp\cdot {\bf b}} \nonumber\\
& = & \frac{1}{4\pi} \nu^2 (\nu b) K_1 (\nu b),
\end{eqnarray}
where $K_n(x)$ is the exponentially-decreasing hyperbolic Bessel
function \cite{19}.

For the $\rho$ meson, we use the density given by the Fourier transform
of the (less-well-known) pion form factor, since it is expected to be
essentially the same as that of the $\rho$ meson in the quark model
\cite{15}. The density $\rho_\rho(b)$ is then given by an expression
similar to that in the first line in Eq.\ (29), but with \cite{17}
\begin{equation}
G_\rho (k_\perp^2)
\approx G_\pi (k_\perp^2)
\approx \left( 1 + \frac{k_\perp^2}{\mu^2}
\right)^{-1},\qquad \mu^2 = 0.47\rm\ GeV^2.
\end{equation}
With this input,
\begin{equation}
\rho_\rho(b) \approx \frac{\mu^2}{2\pi} K_0 (\mu b)
\end{equation}
and \cite{15}
\begin{equation}
A_{\rho p}(b) = \frac{1}{4\pi} \frac{\nu^2\mu^2}{\mu^2 - \nu^2}
\left\{\nu b K_1 (\nu b) - \frac{2\nu^2}{\mu^2 - \nu^2}
\left[ K_0 (\nu b) - K_0 (\mu b)\right] \right\}\ .
\end{equation}

The incident hadronic systems in a $\rho p$ collision can interact
inelastically through soft as well as hard processes, hence the
presence of first term in Eq.\ (24). This soft scattering is dominant
at the energies explored so far. We will parametrize ${\rm Re}\,
\chi_{\rm soft}$ using the same overlap function as above, a choice
which works well in $\bar pp$ and $\pi p$ scattering, and an
energy-dependent factor with a form suggested by Regge theory,
\begin{equation}
2{\rm Re}\,\chi_{\rm soft} = A_{\rho p} (b) \sigma_{\rm soft} (s),
\end{equation}
where \cite{14,20}
\begin{equation}
\sigma_{\rm soft}^{\rho p} (s) \approx \sigma_0 + \sigma_1 (s -
m_p^2)^{-1/2} + \sigma_2(s-m_p^2)^{-1}.
\end{equation}

\subsection{$q\bar q$ - proton cross section}

We will model the $q\bar q$ - proton cross section in Eq.\ (22),
using ideas similar to those above,
but
with changes which reflect the fact that a $q\bar q$ system produced
at an initial momentum transfer $p_{\perp 0}>Q_0 = \mu/2$ does not
have time to develop into a system of normal hadronic size. This
will be reflected in the $q\bar q$ - proton overlap function
$A_{q\bar q \mbox{-} p} (b)$ which will simply approach $\rho_p(b)$
for a pointlike $q\bar q$ system, that is, for $p_{\perp 0}\gg Q_0$.
We will parametrize this change by using a form factor for the $q\bar
q$ system with the same form as $G_\rho$, Eq.\ (30), but with $\mu$
replaced by $2p_{\perp 0}$ as suggested by our earlier discussion,
\begin{equation} G_{q\bar q} (k_\perp^2) = \left( 1 +
\frac{k_\perp^2}{4p_{\perp 0}^2} \right)^{-1},\qquad 4p_{\perp 0}^2
\geq \mu^2 = 0.47\rm\ (GeV/c)^2. \end{equation}
With this choice, the $q\bar q$ profile function $\rho_{q\bar q}(b)$
and the overlap function $A_{q\bar q\mbox{-}p}(b)$ calculated using the
analogs of Eqs.\ (29) and (27) are continuous with $\rho_p(b)$ and
$A_{\rho p}(b)$ at $p_{\perp 0}=Q_0=\mu/2$, and approach a delta
function and $\rho_\rho(b)$ respectively for $p_{\perp 0}\gg Q_0$.

We must also scale $\sigma_{\rm soft}(s)$ in Eq.\ (34) so that the
integral of the eikonal function $2{\rm Re}\,\chi(b)$ (the intrinsic
single-interaction cross section) decreases with the geometrical size
of the $q\bar q$ system for $p_{\perp 0}> Q_0$. We will simply scale by
a factor $(Q_0/p_{\perp 0})^2=(\mu/2p_{\perp 0})^2$ and take
\begin{equation}
\sigma_{\rm soft}^{q\bar qp} = (\mu^2/4p_{\perp 0}^2)\sigma_{\rm
soft}^{\rho p}.
\end{equation}
The $q\bar q$ and vector meson terms are then properly continuous at
$p_{\perp 0} = Q_0$, while the soft $q\bar q$ - proton cross section
decreases as $p_{\perp 0}^{-2}$ for $p_{\perp 0}\gg Q_0$ as is expected
for higher twist contributions. The hard QCD contributions to the
eikonal function for the $q\bar q$ - proton interaction are again of
the form given in Eq.\ (26), but with the parton distributions in
the $\rho$ meson replaced by the parton distributions $f_i^{q\bar
q}(x, p_\perp, p_{\perp 0})$ which evolve from the initial $q\bar
q$ system produced at the transverse momentum $p_{\perp 0}$ and
observed at $p_\perp$.

\subsection{Jet and inclusive cross sections}\label{jet}

The inclusive jet cross section $\sigma_{\rm jet}^{\gamma p}(s,Q^2)$ is
defined to be the part of the inelastic $\gamma p$ cross section which
includes events with at least one semihard parton-parton scattering
with a momentum transfer $p_\perp^2\geq Q^2$, irrespective of any
soft processes which may occur. The semiclassical probability that
there is {\em no} parton-parton scattering
with $p_\perp^2>Q^2$
in a hadronic collision at
impact parameter $b$ is $e^{-2{\rm Re}\,
\chi_{\rm QCD
\rule{0em}{1.25ex}
}(b,s,Q^2)}$. Using this observation, we can rewrite
the cross section
$\sigma_{\rm had}^m$ associated with the hadronic
component $|m\rangle$ of the photon as \cite{10}
\FL\begin{eqnarray}
\sigma_{\rm had}^m &=& \int d^2b\,\left(1-e^{\textstyle-2{\rm Re}\,
\chi^m_{\rm QCD}-2{\rm Re}\,
\chi^m_{\rm soft}}\right)\nonumber\\
&=& \int d^2b\, \left(1-e^{\textstyle -2{\rm
Re}\,\chi^m_{\rm QCD}(b,s,Q^2)}\right)
\nonumber\\
&&+\int d^2b\,\left(1-e^{\textstyle
-2{\rm Re}\,\chi'_{\rm
soft,\ \mit m}(b,s,Q^2)}\right) e^{\textstyle
-2{\rm Re}\,\chi^m_{\rm QCD}
(b,s,Q^2)}\nonumber\\
&=& \sigma^m_{\rm jet}(s,Q^2)+\sigma^m_{\rm
no\mbox{-}jet}(s,Q^2)\ . \label{a42}
\end{eqnarray}
Here
\begin{equation}
\sigma^m_{\rm jet}(s,Q^2)=\int d^2b\,\left(1-e^{\textstyle -2{\rm Re}
\chi^m_{\rm QCD}(b,s,Q^2)}\right)
\label{a43}
\end{equation}
is the total jet cross section. It includes events with multiple
independent parton-parton scatterings (multijet events) as well as
single-scattering events, all with or without accompanying soft
inelastic scattering
processes. The eikonal function ${\rm Re}\chi^m_{\rm
QCD}(b,s,Q^2)$ is defined as in Eqs.\ (25) and (26) with
$p_{\perp,\rm min}^2$ replaced by $Q^2$ in Eq.\ (26).
${\rm Re}\chi_{\rm soft,\ \mit m}'$ in Eq.\ (\ref{a42})
includes both the usual soft term and the contributions from
``soft" jets with $p_{\perp,\rm min}^2\leq p_\perp^2<Q^2$.  It
is given by
the sum of ${\rm Re}\chi^m_{\rm soft}$ and
a modified
${\rm Re}
\chi^m_{\rm QCD}$  calculated
for the restricted interval
 $p_{\perp,\rm min}^2 < p_\perp^2<Q^2$. The cross section
$\sigma^m_{\mbox{\footnotesize no-jet}}$ includes {\em no} jet events
with $p_\perp^2>Q^2$ as is
evident from the
probabilistic interpretation of the
factor $e^{-2{\rm Re}\,\chi_{\rm QCD}^m}$
in its definition
in Eq.\ (\ref{a42}). The total jet cross section in $\gamma p$ scattering
involves direct $\gamma$-parton interactions as well, and
a generalized sum over the hadronic jet cross sections with the
weights ${\cal P}_m$,
\begin{eqnarray}
\sigma_{\rm jet}^{\gamma p} (s, Q^2) &=&
\sigma_{\rm dir}^{\rm QCD} (s, Q^2) +
\sum_m {\cal P}_m \sigma_{\rm jet}^m (s, Q^2)\nonumber\\
&=& \sigma_{\rm dir}^{\rm QCD} (s, Q^2) +
\lambda{\cal P}_\rho \sigma_{\rm jet}^\rho (s, Q^2)\nonumber\\
&& + \sum_q e_q^2 \frac{\alpha_{\rm em}}{\pi}
\int_{Q_0^2} \frac{dp_{\perp 0}^2}{p_{\perp 0}^2}
\sigma_{\rm jet}^{q\bar q p} (s, Q^2, p_{\perp 0}^2).
\label{a45}
\end{eqnarray}

To calculate the multijet cross sections, we note that the average
number of parton-parton scatterings in a collision
at impact parameter $b$ which involves the intermediate hadronic
state $|m\rangle$ is
\begin{eqnarray}
n_m(b,s,Q^2) &=& \sigma_m^{\rm QCD}(s,Q^2)\,A_m(b)\nonumber\\
&=& 2{\rm Re}\,\chi^m_{\rm QCD}(b,s,Q^2)\ .
\label{a46}
\end{eqnarray}
Since the parton scatterings in our model are independent, the
probability of having $n$ scatterings
($2n$ jets) in a hadronic interaction is
given by a Poisson distribution with the average number of scatterings
equal to $n_m(b,s,Q^2)$:
\begin{equation}
P_{n,m}(b,s,Q^2)=\frac{1}{n!}\left[n_m(b,s,Q^2)\right]^n\,
e^{\textstyle -n_m(b,s,Q^2)}\ .
\label{a47}
\end{equation}
The probability distribution for $n=0,1,\ldots$ parton scatterings
or $2n$ jets with $p_\perp^2 \geq Q^2$
averaged over
{\em all} inelastic events is then given by
\begin{eqnarray}
P_n (s, Q^2) &=& \frac{1}{\sigma_{\rm inelas}^{\gamma p} (s) }
\sum_m {\cal P}_m \frac{1}{n!} \int d^2 b
\left[ n_m (b, s, Q^2) \right]^n e^{\textstyle
-n_m(b, s, Q^2)},\ n\geq 1,\label{a48}\\
P_0(s, Q^2) &=& \frac{1}{\sigma_{\rm inelas}^{\gamma p} (s)}
\sum_m {\cal P}_m \int d^2b
\left(1 - e^{\textstyle -2{\rm Re}\,\chi'_{\rm soft,\ \mit m}
(b,s,Q^2)}
\right)\, e^{\textstyle -n_m(b,s,Q^2)}.\ \ \ \
\label{a49}
\end{eqnarray}
The sum in Eq.\ (\ref{a48}) just gives $\sigma_{\rm jet}^{\gamma p}$
up to the very small single-jet contribution from the
direct-interaction cross section in Eq.\ (\ref{a45}). We note that
the factors $e^{-n(b,s,Q^2)}$ and $(1-e^{-2{\rm Re}
\chi_{\rm soft}'})$ in the expression
for $P_0$ are
simply interpreted as
the Poisson probability that there are no hard
parton-parton scatterings, and the probability that there {\em is} a
soft inelastic
interaction in the hadronic collision in question.

We note finally that the average number of parton-parton collisions
with $p_\perp^2 > Q^2$ is given by
\begin{equation}
\bar n (s,Q^2) = \sum_{n=1}^\infty n P_n (s, Q^2).
\label{a50}
\end{equation}
Using the results above and the normalization condition for the
overlap function $A(b)$ in Eq.\ (27), Eq.\ (\ref{a50}) is easily
shown to reduce to the statement that
\begin{eqnarray}
\bar n \sigma_{\rm inelas}^{\gamma p} &=&
\sum_m {\cal P}_m \sigma_{\rm QCD}^m (s, Q^2) \nonumber\\
&=& \sum_{i,j} \int_0^1 dx_1 \int_0^1 dx_2 \int_{Q^2}
dp_\perp^2 f_i^p (x_1, p_\perp^2)
\frac{d\hat\sigma_{ij}}{d p_\perp^2}
\sum_m {\cal P}_m f_i^m (x_2, p_\perp^2).
\label{a51}
\end{eqnarray}
This is just the usual expression for the inclusive hadronic $\gamma
p$ cross section,
\begin{equation}
\bar n\sigma_{\rm inelas}^{\gamma p}
=\sum_{i,j}
\int_0^1 dx_1
\int_0^1 dx_2
\int_{Q^2} dp_\perp^2\,
f_i^p(x_1,p_\perp^2)\,
\frac{d\hat\sigma_{ij}}{dp_\perp^2}\,
f_j^\gamma (x_2,p_\perp^2),
\label{new46}
\end{equation}
but
 with the photon structure functions given
explicitly as weighted sums over the structure functions for the
initial hadronic components of the photon, that is, with
\begin{eqnarray}
f_i^\gamma (x, p_\perp^2) &=& \sum_m {\cal P}_m f_i^m (x, p_\perp^2)
\nonumber\\
& \approx &
\lambda {\cal P}_\rho f_i^\rho (x, p_\perp^2) +
\sum_q e_q^2 \frac{\alpha_{\rm em}}{\pi}
\int_{Q_0^2} \frac{d p_{\perp 0}^2}{p_{\perp 0}^2}
f_i^{q\bar q} (x, p_\perp^2, p_{\perp 0}^2 ).
\label{a52}
\end{eqnarray}
This is just what would be expected for $f_i^\gamma$ in the
standard semiclassical picture of the parton model.

\subsection{Parton distributions in the hadronic photon}

The remaining problem  is the specification of the
parton distributions for the $\rho$ meson and the $q\bar q$ system
which appear in the expression for ${\rm Re}\,\chi_{\rm QCD}$. While
there are no direct measurements of the parton distributions in the
$\rho$, the equivalence of $\rho$ and $\pi$ states in the quark model
up to spin effects suggests that we can equate the distribution
functions $f_i^\rho$ to the corresponding distribution functions
$f_i^\pi$
for
the pion  \cite{21}. Although these are not known as well as
the proton distribution functions, they have been tested in the
present context in an analysis of high-energy $\pi^+ p$ scattering
\cite{15} and seem to be satisfactory.

An alternative possibility which was used in earlier work
\cite{9,11,12} is to equate $f_i^\rho$ to the corresponding
structure function $f_i^\gamma$ for the photon, with the probability
that the photon becomes a vector meson divided out. The
estimate used in \cite{9} and followed in \cite{11} and
\cite{12} was
\begin{equation}
f_i^\rho \approx f_i^\gamma /{\cal P}_\rho.
\label{new48}
\end{equation}
This estimate can be sharpened substantially using Eq.\
(\ref{a52}). It is clearly necessary in this expression to include
the contributions from vector mesons other than the $\rho$, and to
take into account the excited hadronic states which can be
produced at initial momentum transfers
$p_{\perp,0}$ greater than the soft cutoff $Q_0$ but less than
the momentum transfer $p_\perp$ in the parton-level scattering,
that is, in the interval
$Q_0<p_{\perp0}<p_\perp$.

The soft contributions to $f_i^\gamma$ associated with excited
vector states are expected to fall as $1/p_{\perp0}^2$, but can be
significant. We can estimate these contributions by assuming that
\begin{equation}
f_i^{q\bar q}(x,p_\perp^2,p_{\perp0}^2)
\approx
\left( \frac{Q_0}{p_{\perp0}}\right)^2
f_i^\rho (x,p_\perp^2)
\label{new49}
\end{equation}
for
$p_\perp>
p_{\perp0}>Q_0$ (again enforcing continuity in the
distributions) and integrating Eq.\ (\ref{a52}).  The result for
three light quarks
and $p_\perp\gg Q_0$
is
\begin{equation}
f_i^\gamma \approx \left(\lambda{\cal P}_{\rho} +
\frac{2}{3}\frac{\alpha_{em}}{\pi}\right) f_i^\rho.
\label{new50}
\end{equation}
The effect is to replace ${\cal P}_\rho$ in Eq.\ (\ref{new48}) by
\begin{equation}
{\cal P}' = \lambda {\cal P}_\rho
\left( 1 + \frac{2}{3\pi\lambda} \frac{f_\rho^2}{4\pi}\right).
\label{new51}
\end{equation}
Numerically, ${\cal P}' \approx 1.8 {\cal P}_\rho$ for
$f_\rho^2/4\pi = 2.2$ and $\lambda=4/3$ (equal $Vp$ cross
sections for the $\rho,\omega$ and $\phi$ mesons), and ${\cal P}'
\approx 1.58{\cal P}_\rho$ for $\lambda=10/9$ (complete
suppression of the $\phi$ contribution).

Additional
purely perturbative contributions to the $f_i^\gamma$ are generated
by the inhomogeneous term in the QCD evolution equations for the quark
distributions in the photon \cite{22},
\begin{eqnarray}
\frac{dq_i^\gamma}{d \ln(Q^2/\Lambda^2)}(x,Q^2)
&=& 3e_i^2\frac{\alpha_{\rm em}}{2\pi}\left[x^2 + (1-x)^2\right]
\nonumber\\
&& + \frac{\alpha_s}{2\pi}
(Q^2)
\int_x^1 \frac{dy}{y}
\left[P_{qq}\left(\frac{x}{y}\right) q_i^\gamma (y,Q^2)+
P_{qG} \left(\frac{x}{y}\right) G^\gamma (y,Q^2)\right]\, ,\nonumber \\
\frac{dG^\gamma }{d\ln(Q^2/\Lambda ^2)}(x,Q^2) &=&
\frac{\alpha_s(Q^2)}{2\pi
}\int_{x}^{1}\frac{dy}{y}\left[ \sum_{i=1}^{2f}P_{Gq}\left( \frac{x}{y}
\right)q_{i}^{\gamma }(y,Q^2)\right.\nonumber \\
&&\left.+P_{GG}\left( \frac{x}{y} \right)G^{\gamma
}(y,Q^2) \right]
\end{eqnarray}
where the $P$'s are the usual quark and gluon splitting functions.
Forshaw \cite{new18}  estimated these
``pointlike" contributions starting
with vanishing distribution functions
at
$p_{\perp 0} = 1$ GeV, and found them to be small
at the momentum transfers $p_\perp$ which are relevant for our
considerations.

Finally, for either of the foregoing models for $f_i^\rho$, we can
estimate $f_i^{q\bar q}\left( x,p_\perp^2,p_{\perp0}^2 \right)$
simply as $\left(Q_0/p_{\perp 0}\right)^2
f_i^\rho\left( x,p_\perp^2 \right)$. This estimate assumes that the
initial nonperturbative
parton distributions $f_i^{q\bar q}$ at
$p_{\perp 0}>Q_0$ can be equated to the $\rho$ meson distributions
at $p_{\perp0}$ scaled by the size of the system,
with the distribution then evolved to
$p_\perp$ using the homogeneous part of the QCD evolution
equations
(the evolution equations for the $\rho $ or $\pi $ do not have an
inhomogeneous term).  The final distributions therefore include the
effects of evolution all the way from $Q_0$ to $p_{\perp}$. A better
approximation would presumably be to use the scaled distributions
at $Q_0$ as input for evolution from $p_{\perp0}$ to $p_\perp$,
again using the homogeneous version of the evolution equations. This
calculation may become worthwhile in the future, but we do not
expect the differences from the present estimates to be large
since large values of
$p_\perp$ and $p_{\perp0}$ are suppressed by the rapid
decrease in the parton-level cross sections and the
$(Q_0/p_{\perp0})^2$ scaling, respectively.

To summarize, the total inelastic $\gamma p$ cross section is given by
\begin{eqnarray}
\sigma_{\rm inel}^{\gamma p} &=& \sigma_{\rm dir} +
\lambda {\cal P}_p \int d^2 b \left( 1 - e^{-2{\rm Re}\,\chi_{\rho
p}}\right) \nonumber\\
&& + \sum_q e_q^2 \frac{\alpha_{\rm em}}{\pi}
\int_{Q_0^2} \frac{dp_{\perp 0}^2}{p_{\perp 0}^2}
\int d^2 b
\left( 1 - e^{-2{\rm Re}\,\chi_{q\bar qp}}\right)\ .
\label{new54}
\end{eqnarray}
Here
\begin{equation}
2{\rm Re}\,\chi_{\rho p}(b,s) = A_{\rho p}(b)
\left[ \sigma_{\rm soft}(s) + \sigma_{\rm QCD}^{\rho p}(s) \right]
\label{new55}
\end{equation}
where $A_{\rho p}$, $\sigma^{\mit \rho p}_{\rm soft}$, and
$\sigma_{\rm QCD}^{\rho p}$ are given in Eqs.\ (32), (34), and (26),
respectively. Finally,
\begin{equation}
2{\rm Re}\,\chi_{q\bar qp} (b,s,p_{\perp 0}) = A_{q\bar qp}
(b,p_{\perp 0}) \left[ \sigma_{\rm soft}^{q\bar qp}
(s, p_{\perp 0}) + \sigma_{\rm QCD}^{q\bar qp}
(s,p_{\perp 0}) \right],\label{new56}
\end{equation}
where $A_{q\bar qp}$ is given by Eq.\ (32) with $\mu$ replaced
by $2p_{\perp 0}$, $\sigma_{\rm soft}^{q\bar qp}$ is given by Eq.\
(36), and $\sigma_{\rm QCD}^{q\bar qp}$ is given by the
analog of Eq.\ (26) with $f_i^\rho$ replaced by $f_i^{q\bar q}
\approx (Q_0/ p_{\perp 0} )^2 f_i^\rho$.

The results above differ from those in earlier work in important
ways. Thus the authors of \cite{9}, \cite{11}, and \cite{12} omit
the last term in Eq.\ (\ref{new54}), estimate the $\rho $ meson
structure functions $f_i^\rho $ in $\chi _{\rho p }$ using the
approximation in Eq.\ (\ref{new48}), and ignore contributions other
than those of the $\rho $, i.e., take $\lambda =1$ in Eq.\
(\ref{new54}). Forshaw's recent analysis \cite{new18} includes the last
term in Eq.\ (\ref{new54}), but with $\chi _{q\bar qp}$ restricted
to the purely ``pointlike" contributions which he shows are small.
His discussion neglects the residual soft contributions from excited
states. These fall as $1/p_{\perp0}^2$ in our model as expected for
``higher twist" contributions, but are nevertheless important. The
last term contributes approximately 1/3 of the cross section at the
energies of interest in our applications. The direct cross section
is quite small, roughly 1.5\%\ of the total at HERA energies. We
re\,emphasize, finally, that continuity in the expressions at
$p_{\perp0}=Q_0$ is essential since ``hard" and ``soft" processes
are not clearly separated and must merge smoothly.

\section{RESULTS FOR THE
$\gamma \mbox{\lowercase{$p$}}$ and $\gamma$-NUCLEUS CROSS
SECTIONS}

\subsection{$\gamma p$ cross sections}

Our calculations of the inelastic $\gamma p$ cross sections were
based on Eq.\ (\ref{new54}
), with the eikonal functions parametrized
as described earlier. We  note, in particular, that our
Regge-type parametrization of $\sigma_{\rm soft}$ allows us to
connect the high- and low-energy cross sections smoothly. Previous
calculations \cite{3,6,11,12} have taken $\sigma_{\rm soft}$ as
constant, with the value chosen to fit the measured $\gamma p$
cross section near its minimum at $\sqrt{s}\approx8$ GeV,
and fail at lower energies.
The (eikonalized) soft cross section
in the ``high-energy" range 7 GeV $\leq\sqrt{s}\leq$ 20 GeV
actually involves a significant
contribution  from the decreasing terms in Eq.\
(34) which are necessary to fit the lower-energy data.
This affects the value of $p_{\perp,\rm min}$
needed to fit these data
using the
rising QCD contributions,
hence affects the
increase in the cross section predicted
at high energies. We obtain quite
reasonable fits to the data for 3 GeV $\leq$ 10 GeV \cite{24}
using the soft cross section in
Eq.\ (34), but have not tried to obtain a ``best" fit given
the apparent inconsistencies in normalization between different
experiments.

The cross section $\sigma_{\rm had}^{QCD}$ for semihard scattering
was calculated using the exact formulas for the O($\alpha_s^2$)
parton-parton cross sections. The strong coupling $\alpha_s(Q^2)$
was evaluated at the scale $Q^2=p_\perp^2$ for four flavors with
$\Lambda_{QCD}=200$ MeV. We used the proton structure functions of
Eichten, Hinchliffe, Lane, and Quigg \cite{25,26}. The results do
not change significantly for different parametrizations \cite{18}
of the proton structure functions \cite{6}.
For the two models we considered for the $\rho$ meson and $q\bar q$
structure functions, we used the pion structure functions of Owens
\cite{21} and the photon
structure functions of Drees and Grassie \cite{27}. The photon
structure functions of Duke and Owens \cite{28} give very similar
results in the region $\sqrt{s}\alt20$ GeV in which there are data,
but are unrealistically singular for $x\rightarrow0$, and are not
reliable for high-energy applications \cite{6}.
The direct cross
section in Eq.\ (7) was also calculated using the proton
structure functions of Eichten {\em et al.} \cite{25}, and the same
value of $p_{\perp,\rm min}$
as used in the hadronic terms. The
direct cross section is always very small relative to the hadronic
cross section.

For the model based on the pion structure functions, we used the
value $p_{\perp,\rm min}=1.45$ GeV which was fixed in earlier fits
to the $\pi p$ scattering cross sections \cite{15}. The only
adjustable parameters were then the three cross sections
$\sigma_0,\sigma_1,\sigma_2$ which appear in the soft cross
section in Eq.\ (34), and the parameter $\lambda$ which specifies
the extent to which the $\phi$ meson contributes to the sum of
cross sections in Eq.\ (10). The parameter $\sigma_2$ is important
only at very low energies. The remaining parameters are reasonably
well fixed by the data between $\sqrt{s}=3$ GeV and $\sqrt{s}=20$
GeV once $\lambda$ is specified. The parameters used for the two
cases considered, $\lambda=4/3$ (no $\phi$ suppression)
and $\lambda=10/9$ (complete $\phi$ suppression) are
listed in Table I.

The calculated cross sections are compared to the data below
$\sqrt{s}=20$ GeV in Fig.\ 1 which shows
how the use of a Reggeized
soft cross section gives a smooth connection between the low-energy
region 3 GeV $\leq\sqrt{s}\leq10$ GeV and the high-energy region
10 GeV $\leq\sqrt{s}\leq20$ GeV. The effects of semihard parton
scatterings begin to be important in the latter region, and account
for the observed increase in the inelastic $\gamma p$ scattering
cross section. We have not attempted to fit the cross section data
below $\sqrt{s}=2$ GeV ($E_{\gamma,\rm lab}=1.66$ GeV) where
resonance effects become important.

We show the results for the region 3 GeV $\leq\sqrt{s}\leq$ 20 GeV
in more detail in Fig.\ 2. This figure clearly shows the increase
in $\sigma_{\rm inel}^{\gamma p}$ associated with the onset of
semihard parton scatterings. It shows also the importance of
including the lower-energy region in the analysis: the observed
rise in the cross section is relative to a falling contribution
from soft processes in this energy region. This ``background" is
shown in Fig.\ 2 as the curve for the eikonalized soft cross
section calculated including the soft contributions from the last
two terms in Eq.\ (\ref{new54}),
plus the very small contribution from the direct cross section.
The two cases considered,
$\lambda=4/3$ and $\lambda=10/9$, give results
for this background
which are practically indistinguishable  even though
the soft backgrounds are somewhat different.

The cross sections predicted for the two cases diverge mildly at
higher energies as shown in Fig.\ 3. The divergence is primarily
due to the suppression of the $\phi$ contribution for
$\lambda=10/9$. The soft background in the two cases
differs by less than 0.001 mb at $\sqrt{s}=400$ GeV. A
curve close to the
lower
curve with $\lambda=10/9$ is probably to be preferred
since the data on photoproduction for
$\sqrt{s}\approx
10$ GeV show strong suppression of
the $\phi/\rho^0$ production ratio
relative to
the expectation for the vector dominance model with quark
model couplings \cite{16,29}.  In addition, the ``$\pi p$" cross
section calculated using Eq.\ (23) and the soft and     QCD
contributions for $\lambda=10/9$ is rather close to the
actual $\pi N$ cross section as it should be in a model which
equates $\rho p$ and $\pi p$ scattering. (The ``$\pi p$" cross
section is
about 11\% low in this case
at $\sqrt{s}=10$ GeV, and about 21\% low when calculated
using the soft
background
for $\lambda=4/3$.)

As shown in Fig.\ 3 the predictions for $\sigma_{\rm inel}^{\gamma
p}$ based on the pion structure functions are
consistent with, but perhaps
somewhat higher than,
the preliminary measurements at HERA reported by the ZEUS and H1
collaborations \cite{30}. Since the main contribution to
$\sigma_{\rm QCD}$ at high energies arises from the scattering of
low-$x$ gluons,
a high value of the
calculated
cross section would
 suggest that the gluon content
of the pion had been overestimated in \cite{21}.
The approximate photon structure functions given by the
expression in Eqs.\ (\ref{a52}) and (\ref{new50}) with
$f_i^\rho$ equated to $f_i^\pi$ would then have  a gluon
content which is also somewhat too large
at low $x$.

Our results for the model based on the photon structure functions of
Drees and Grassie \cite{27} with the factor ${\cal P}'$ in Eq.\
(\ref{new51}) divided out are shown in Figs.\ 4 and 5. The curves
shown in these figures were calculated using $p_{\perp,\rm min}=1.2$
GeV and 1.4 GeV and the soft parameters given in Table \mbox{I}. It
was assumed that the $\phi$ contribution is suppressed, though this
makes little difference in the final results. The curves shown in
Fig.\ 4 bracket the data for 6 GeV $\leq\sqrt{s}\leq20$ GeV. An
intermediate value of the QCD cutoff, $p_{\perp,\rm min}\approx3$
GeV, is clearly favored. The results obtained at higher energies
with the Drees-Grassie structure functions are shown in Fig.\ 5. The
calculated cross sections are significantly higher than those
obtained using the pion structure functions, and rise more rapidly
at high energies. The predictions at $\sqrt{s}=200$ GeV are clearly
inconsistent with the preliminary measurements of $\sigma_{\rm
inel}^{\gamma p}$ at HERA \cite{30}. The cutoff $p_{\perp,\rm min}$
would have to be increased to roughly 1.7 GeV to bring the
calculated cross section into agreement with the HERA data, but such
a large value of $p_{\perp,\rm min}$ would be inconsistent with the
data for $\sqrt{s}<20$ GeV as shown by Fig.\ 4. We therefore
conclude that these photon structure functions are not satisfactory
in the present application. The very rapid rise in the calculated
cross sections for $\sqrt{s}\agt 200$ GeV is associated with the
rapid growth of the parton distribution functions at the small
values of $x$ which become accessible at these energies.

\subsection{$\gamma$-air cross sections}

Our model for the hadronic interactions of the photon can be
extended to photon-nucleus interactions using Glauber's multiple
scattering theory \cite{31} with the result
\begin{eqnarray}
\sigma_{\rm inel}^{\gamma\mbox{-}\rm air}
&=& A \sigma_{\rm dir} + \lambda{\cal P}_\rho
\int d^2 b \,\langle \Psi| 1 -
{\rm exp}\left(-\sum\nolimits_{j=1}^A2{\rm Re}\,\chi_j^{\rho p}
\right)
|\Psi\rangle\nonumber\\
&& +\sum_q e_q^2\,\frac{\alpha_{em}}{\pi}
\int_{Q_0^2} \frac{dp_{\perp0}^2}{p_{\perp0}^2}
\int d^2 b\,\langle \Psi |\,
1-{\rm exp}\left( -\sum\nolimits_{j=1}^A 2{\rm Re}\chi_j^{q\bar qp}
\right) |\Psi \rangle.
\label{new57}
\end{eqnarray}
The expectation values are to be taken in the nuclear ground
state. ${\rm Re}\chi_j^m={\rm Re}\chi^m(|{\bf b}-{\bf r}_{j
\perp}|)$ is the eikonal function for the scattering of the
hadronic component $|m\rangle$ of the photon on the $j^{\rm th}$
nucleon, where  ${\bf r}_{j\perp}$ is the instantaneous
transverse distance of that nucleon from the nuclear center of
mass, and $\bf b$ is the impact parameter of the photon relative
to the nucleus. The integrals were evaluated using shell-model
wave functions for the oxygen and nitrogen nuclei \cite{32}. The
eikonal functions for $\gamma n$ and $\gamma p$ scattering were
taken as equal.

In Fig.\ 6 we show the inelastic $\gamma p$ cross section for
laboratory photon energies up to $2\times10^8$ GeV
($\sqrt{s}\approx2\times10^4$ GeV), calculated as described above
using the parton distribution functions of Eichten {\em et al.}
\cite{25} for the proton and the pion structure functions of Owens
\cite{21} for the photon
with the $\phi$ contribution suppressed.
The eikonalized $\gamma p$ cross section
rises to 0.5 mb at $E_{\gamma,\rm lab}=2\times10^8$ GeV, a large
value for a photon-initiated process, but still a small cross
section on a hadronic scale.
Since the results obtained in \cite{15} for $\pi p$ scattering at
equivalent energies
suggest that the gluon content of the pion given by \cite{21}
is too large, the $\gamma p$ cross section shown in Fig.\ 6 should
be regarded as an upper limit.
A generous upper bound on the $\gamma$-air
cross section is given by $A\sigma_{\rm inel}^{\gamma p}$ which
reaches 7.2 mb for $E_{\gamma,\rm lab}=2\times10^8$ GeV and
$A_{\rm air}\approx 14.4$. The actual cross section
is lower, as shown in Fig.\ 6, reaching only 3 mb at
$E_{\gamma,\rm lab}=2\times10^8$ GeV.
The limit is far below the result
$\sigma_{\rm inel}^{\gamma p}\approx 90$ mb obtained in \cite{6}
using the incorrect eikonalization procedure noted in the
Introduction, and is clearly far too small to account for the
reported muon anomalies in cosmic ray air showers.
The $\gamma$-air cross section may still be
large enough to be interesting for shower evolution.

\subsection{Conclusions and comments}

Our main conclusion is simple: the proper treatment of
eikonalization in the calculation of inelastic $\gamma p$
interactions is important even at rather low energies. This is in
accord with the observations of Collins and Ladinsky \cite{9}.
However, the situation is somewhat more complex than envisaged by
those authors. The $\gamma p$ cross section is given as a weighted
sum of cross sections for the different hadronic components of the
photon, and cannot be expressed directly in terms of the photon
structure functions.

We have formulated a simple model which includes contributions to
the cross section from the low-mass vector mesons and from excited
states described in a quark-gluon basis. The soft cross sections and
quark and gluon distributions for the excited states are taken as
scaled versions of the corresponding quantities for the $\rho$
meson, and the distribution functions for the $\rho$ are equated to
those for the pion. The model works well at lower energies using the
cutoff $p_{\perp,\rm min}$ in the QCD contributions which was
determined for $\pi p$ scattering in \cite{15}, and gives a
reasonable but perhaps somewhat high cross section at the energy of
the preliminary measurements at HERA, $\sqrt{s}\approx200$ GeV
\cite{30}. The discrepancy, if real, suggests that the gluon content
of the pion has been overestimated at small $x$ in \cite{21}, a
conclusion consistent with that in \cite{15}.

We emphasize that our result at $\sqrt{s}=200$ GeV is a genuine
prediction. The parametrization of the low-energy cross section is
essentially independent of the very small QCD contributions in
that energy range. The only remaining parameter, $p_{\perp,\rm
min}$, was fixed from the $\pi N$ scattering data, and not
adjusted to fit either the rise in $\sigma ^{\gamma p}$ observed
for 8 GeV $\alt\sqrt{s}\alt20$ GeV, or the HERA data.

The model also gives an interesting {\em a priori} estimate for
the quark and gluon distributions in the photon,
\begin{equation}
f_i^\gamma = \lambda \frac{4\pi\alpha_{em}}{f_\rho^2}
\left( 1+\frac{2}{3\pi\lambda}\frac{f_\rho^2}{4\pi}\right)
f_i^\pi,
\label{new58}
\end{equation}
where $10/9 < \lambda < 4/3$. It will be interesting to compare
future direct measurements of $f_i^\gamma$ at HERA with this
estimate.

The version of the model in which the $\rho$ distribution functions
are estimated from the photon structure functions $f_i^\gamma$ gives
cross sections which are too large to agree with the new HERA data
\cite{30} if the Drees-Grassie functions $f_i^\gamma$ are used in
that calculation. We
have not checked other sets of photon structure functions
directly. However, recent calculations by Forshaw and Storrow
\cite{fst33} using the photon distribution functions of Gl\"uck,
Reya, and V\"ogt \cite{34} and Abramowicz, Charchula, and Levy
\cite{35}, give larger values of $\sigma _{\rm QCD}^{\gamma p}$ at
$\sqrt{s}=200$ GeV than those obtained with the Drees-Grassie
distribution functions \cite{27}, so a change to those $f^{\gamma
}$'s would not improve the fits.

It should be emphasized in this connection that the requirement that
a QCD-based model fit the $\gamma p$ cross section in the lower
energy region 5 GeV $\alt\sqrt{s}\leq20$ GeV is a nontrivial
constraint. The data through this region display upward curvature
around a minimum near 8 GeV as is evident in Fig.\ 2. The decrease
in the cross section as $\sqrt{s}$ increases toward the minimum
indicates the presence of a falling non-perturbative background, and
the perturbative QCD contributions are to be determined relative to
this background whatever the details of the model. This restricts
the choice of $p_{\perp,\rm min}$ to values lower than those
characteristic of fits with a constant background \cite{6,9,11,12},
hence influences the predictions for $\sigma ^{\gamma p}$ at high
energies. The constant-background fits are very poor for
$\sqrt{s}<8$ GeV.

Finally, extrapolation of $\sigma_{\rm inelas}^{\gamma p}$ and
$\sigma_{\rm inelas}^{\gamma\mbox{-}\rm air}$ to the highest
cosmic ray energies gives cross sections of a few millibarns,
large enough to be interesting, but much too small to account for
the reported muon anomalies in photon-initiated air showers. The
ultra-high-energy cross sections are of course very uncertain
because of our lack of detailed knowledge of the small-$x$
behavior of the parton distributions in the $\rho$ meson and
nucleon, but the effects of proper eikonalization are strong
enough that much larger cross sections would appear to be
precluded.

\acknowledgements

This work was supported in part through U.S. Department of Energy
Grants Nos. DE-AC02-76ER00881 and DE-FG02-85ER40213, and in part
by the World Laboratory. One of the
authors (LD) would like to thank Dr.\ J.R. Forshaw for useful
correspondence, and the Aspen Center for Physics for its
hospitality while parts of this work were done.

\figure{The inelastic $\gamma p$ scattering cross section for
$\sqrt{s}\leq20$ GeV predicted by the model in the text using pion
structure functions. The data are from the references in
\cite{24}. The soft contributions to the cross section were
determined using data for 3 GeV $\leq\sqrt{s}\leq8$ GeV only. The
QCD contributions were calculated using $p_{\perp,\rm min}=
1.45$ GeV, the value determined in \cite{15} in a fit to $\pi^\pm
p$ cross sections for 10 GeV $\leq\sqrt{s}\leq26$ GeV.}

\figure{The inelastic $\gamma p$ cross section for
3 GeV $\leq\sqrt{s}\leq20$ GeV compared to the cross sections
predicted using pion structure functions, with the $\phi$ meson
contribution at full
strength or totally suppressed. The data shown are
from \cite{24}.}

\figure{The predictions for $\sigma_{\rm inel}^{\gamma p}$ for
$\sqrt{s}\leq400$ GeV for the model based on pion structure
functions with the $\phi$ meson contribution present or totally
suppressed. The lower-energy data are from \cite{24}. The
preliminary HERA data at $\sqrt{s}\approx200$ GeV are from
\cite{30}.}

\figure{The inelastic $\gamma p$ cross sections for
3 GeV $\leq\sqrt{s}\leq20$ GeV compared to the cross sections
predicted using the Drees-Grassie structure functions for the
photon with $p_{\perp,\rm min}=1.2$ GeV and 1.4 GeV and complete
suppression of the $\phi$ meson contribution. The data are from
\cite{24}.}

\figure{The predictions for $\sigma_{\rm inel}^{\gamma p}$ for
$\sqrt{s}\leq400$ GeV for the model base on the Drees-Grassie
structure functions for the photon with $p_{\perp,\rm min}=
1.2$ GeV and 1.4 GeV, and complete suppression of the $\phi$ meson
contribution.
The lower-energy data are from \cite{24}. The
preliminary HERA data at $\sqrt{s}\approx200$ GeV are from
\cite{30}.}

\figure{The predictions for $\sigma_{\rm inel}^{\gamma p}$ and
$\sigma_{\rm inel}^{\gamma\mbox{-}\rm air}$ for $\sqrt{s}
\leq 2\times10^4$ GeV or $E_{\gamma,\rm lab}\leq 2\times10^8$ GeV
for the model based on pion structure functions.}

\begin{table}
\caption{Values of the parameters used in the calculations shown
in Figs.\ 1--5.}
\begin{tabular}{ccccc}
Structure functions & $\sigma_0$ (mb) & $\sigma_1$ (mb GeV)
& $\sigma_2$ (mb GeV$^2$) & $p_{\perp,\rm min}$ (GeV)\\
\tableline
Owens (pion) & 25.8 & 14.0 & 14.0 & 1.45\\[-2ex]
$\phi$ suppressed &&&&\\[1ex]
Owens (pion) & 22.0 & 11.7 & 11.7 & 1.45\\[-2ex]
$\phi$ included &&&&\\[1ex]
Drees-Grassie & 25.8 & 14.0 & 14.0 & 1.2, 1.4\\[-2ex]
$\phi$ suppressed &&&&
\end{tabular}
\end{table}


\begin{references}
\bibitem{13}
We would like to thank J.R. Forshaw for pointing out an important error
in our first attempt at this extended calculation.
\bibitem{30}
The preliminary values of $\sigma_{\rm inel}^{\gamma p}$ reported
at the XXVI International Conference on High Energy Physics were:
ZEUS Collaboration (reported by B. L\"ohr) $\sigma^{\gamma p}=
153\pm 16 \pm 35\ \mu\rm b$ at an average energy $\sqrt{s}=210$
GeV; H1 Collaboration (reported by F. Eisele), $\sigma_{\gamma p}
=165 \pm 28\ \mu\rm b$ or $150 \pm 22\ \mu\rm b$ at $\sqrt{s}
=210$ GeV, depending on the model used in the analysis. We show a
value $160 \pm 30\ \mu\rm b$ for H1 in our figures to span the
theoretical and experimental uncertainties.
\bibitem{20}
The proper variable for the Regge expansion is $s-m_\rho^2$ and not
$s$. The former is proportional to $\cos\theta_t$, the cosine of the
$t$-channel scattering angle. It is also proportional to $E_{\gamma,\rm
lab}$, the variable used in the expansions reported in [14].
\bibitem{29}
We pick this energy for the comparison because it is clearly in
the flat region near the minimum in $\sigma^{\gamma\rho}$, but at
an energy at which $s\bar s$ contributions to the cross section
which we treat at full strength are not yet large. The suppression
is greater at lower energies.
\end{references}
\end{document}